\documentclass[12pt]{article}

\usepackage{epsf}
\usepackage{rotating}

\def\text{\textstyle}

\def\tgb{\mbox{tg}\beta}

\newcommand{\lsim}{\raisebox{-0.13cm}{~\shortstack{$<$ \\[-0.07cm] $\sim$}}~}

\newcommand{\TeV}{\unskip\,\mathrm{TeV}}
\newcommand{\GeV}{\unskip\,\mathrm{GeV}}

\newcommand{\fb}{\unskip\,\mathrm{fb}}

\catcode`\@=11
\def\citer{\@ifnextchar [{\@tempswatrue\@citexr}{\@tempswafalse\@citexr[]}}
\def\@citexr[#1]#2{\if@filesw\immediate\write\@auxout{\string\citation{#2}}\fi
  \def\@citea{}\@cite{\@for\@citeb:=#2\do
    {\@citea\def\@citea{--\penalty\@m}\@ifundefined
       {b@\@citeb}{{\bf ?}\@warning
       {Citation `\@citeb' on page \thepage \space undefined}}%
\hbox{\csname b@\@citeb\endcsname}}}{#1}}
\catcode`\@=12

\begin{document}

\renewcommand{\thefootnote}{\fnsymbol{footnote}}
\setcounter{page}{0}

\begin{titlepage}

\begin{flushright}
hep-ph/0101314
\end{flushright}

\vspace*{1.0cm}

\begin{center}
{\large \sc Higgs Radiation off Quarks \\[0.3cm]
in the Standard Model and supersymmetric Theories \\[0.3cm]
at $e^+e^-$ Colliders}\\
\end{center}

\vskip 1.cm
\begin{center}
{\sc Michael Spira}

\vskip 0.8cm

\begin{small} 
{\it Paul--Scherrer-Institut, CH--5232 Villigen PSI, Switzerland}
\end{small}
\end{center}

\vskip 2cm

\begin{abstract}
\noindent
Yukawa couplings between Higgs bosons and quarks in the Standard Model (SM)
and supersymmetric theories can be measured in the processes $e^+e^-\to
Q\bar{Q} + \mbox{Higgs}$. The cross sections and Higgs energy distributions
of these processes in the SM and minimal supersymmetric model have been
determined including the complete set of next-to-leading order QCD
corrections for all channels.
\end{abstract}

\vfill
\begin{center}
Presented at the \\[1cm]
\begin{large}
5th International Symposium on Radiative Corrections \\
(RADCOR--2000) \\[4pt]
Carmel CA, USA, 11--15 September, 2000
\end{large}
\end{center}
\vfill

\end{titlepage}

\renewcommand{\thefootnote}{\arabic{footnote}}

\setcounter{footnote}{0}

\section{Introduction}
In the Standard Model (SM) and its supersymmetric extensions, the
masses of electroweak gauge bosons, leptons, and quarks are generated
by interactions with Higgs fields \cite{hi64}. The Yukawa couplings
between Higgs particles and fermions therefore grow with the masses
$M_f$ of the fermions. The couplings obey a universal scaling law
$g_{ffH}=M_f/v$ in the SM, with $v\approx 246\GeV$ being the
ground-state value of the Higgs field. In supersymmetric models, which
involve at least two Higgs doublets, the size of the Yukawa couplings
is also set by the fermion masses, yet the relationship is more
complex owing to the mixing among the Higgs fields. The Yukawa couplings
\cite{gu86} of the two CP-even light/heavy Higgs bosons $h/H$
and of the CP-odd Higgs boson $A$ in the minimal supersymmetric
extension of the Standard Model (MSSM) \cite{mssma,mssmb}, expressed in
units of the SM couplings, are determined by
the parameters $\tgb=v_2/v_1$, the ratio of the vacuum expectation values
of the two Higgs fields generating the masses of up- and down-type
particles, and $\alpha$, the mixing angle in the CP-even sector. In
the decoupling limit, in which the light Higgs mass reaches the
maximum value for a given parameter $\tgb$, the $h$ Yukawa couplings
approach the SM values.  In
general, they are suppressed for up-type fermions and enhanced for
down-type fermions; the enhancement increases with $\tgb$ and can
therefore be very strong.

Higgs radiation off top 
or off bottom quarks in $e^+e^-$ collisions,
\begin{equation}
e^+e^- \to Q\bar{Q} \phi \qquad [Q=t,b;\; \phi=H_{SM},h,H,A],
\label{eq:processes}
\end{equation}
lends itself as a suitable process for measuring the Yukawa couplings
of all Higgs bosons \cite{dj92}, particularly for the SM Higgs bosons,
the light Higgs boson $h$ and for moderately heavy Higgs bosons $H$ and $A$.
Within the SM it has been demonstrated that the top Yukawa coupling can
be measured with an accuracy of 2--3\% at linear colliders with
$\sqrt{s}=800$ GeV \cite{tyukawa}.
We present the cross sections for these processes
including the next-to-leading order (NLO) QCD corrections
\cite{di98,di00}.
In this analysis we have calculated the complete set of ${\cal
O}(\alpha_\mathrm{s})$ QCD corrections to all subchannels in
(\ref{eq:processes}) systematically. The large number of interfering 
subchannels in supersymmetric theories renders this program 
more complex than the corresponding calculation in the SM, in
particular since the relative weight of the subchannels varies over
the supersymmetric parameter space and over the phase space for
different mass ratios. Moreover, the Higgs-energy distributions haven
been obtained at NLO \cite{dixx}. They turn out to be relevant in the
separation of the resonant parts and those, which depend on the Yukawa
couplings. Introducing appropriate cuts in the Higgs energy the sensitivity
to the Yukawa couplings can be increased experimentally.

\section{QCD Corrections}
The QCD corrections can be categorized into five classes. Virtual corrections
of the internal quark lines, of the $\gamma/Z$--quark vertices, of the
Higgs--quark vertices,
and box diagrams interfere with the Born amplitude. Gluon radiation
off internal and external quark lines adds incoherently to the cross
sections.
The value of the electromagnetic coupling is taken to be $\alpha =
1/128$ and the Weinberg angle as $\sin^2 \theta_W = 0.23$.
The mass of the $Z$~boson is set to $M_Z=91.187\GeV$, and the
pole masses of the top and bottom quarks are set to $M_t=174\GeV$
\cite{ca98}
and\footnote{This value for the perturbative pole mass of the
bottom quark corresponds in NLO to an $\overline{\rm MS}$ mass
$\overline{m}_b(\overline{m}_b)=4.28\GeV$.}
$M_b=4.62\GeV$ \cite{bottom}, respectively. The masses of the MSSM Higgs
bosons and their couplings are related to
$\tgb$ and the pseudoscalar Higgs boson mass $M_A$. In
the relation we use, higher-order corrections up to two loops
in the effective-potential approach are included \cite{ca95}.
The SUSY parameters are chosen as $\mu=A_t=A_b=0$ and $M_{\tilde{Q}} =
1$ TeV; this simple choice is sufficient to illustrate the main results.

The production of $b\bar{b}\phi$ final states can be mediated by resonance
channels $e^+e^-\to ZH_{SM},Zh,ZH$ and $e^+e^-\to Ah,AH$. We describe
the resonance structures as Breit--Wigner forms by
substituting $M^2\to M^2-iM\Gamma$ in all boson propagators.
The decay widths of the Higgs bosons are
calculated including higher-order corrections, as described in 
Refs.~\cite{12a,dj95}, while the $Z$ width is set to $\Gamma_Z=2.49\GeV$.
For $t\bar{t}\phi$ production the widths can be
neglected, since the Higgs masses are taken below the 
$t\bar t$ threshold.
The renormalization scale of the QCD coupling $\alpha_{\mathrm{s}}$,
which is evaluated in NLO with five active flavors normalized to
$\alpha_{\mathrm{s}}(M_Z^2) = 0.119$ \cite{ca98},
is chosen at $\mu_R^2=s$, where $s=E_{\mathrm{CM}}^2$
is the center-of-mass (CM) energy squared.

The QCD radiative corrections have been calculated in the standard
way. The Feynman diagrams have been evaluated within dimensional
regularization. Ultraviolet divergences are consistently regularized in
$D=4-2\epsilon$ dimensions, with $\gamma_5$ treated naively since no
anomalies are involved.  The renormalization of the 
$Q\bar{Q}\phi$ vertices is
connected to the renormalization of the quark masses, which, in the case
of the top quark, is defined on shell (see e.g.\ Refs.~\cite{de93,bl80}
for details). In the case of bottom quarks large logarithms
are mapped into the running mass $\overline{m}_b(Q^2_{\mathrm{Higgs}})$ for the
Yukawa couplings of the $b$~quark, with $Q^2_{\mathrm{Higgs}}$ denoting
the squared momentum flow through the corresponding Higgs-boson line.
The infrared divergences encountered in the virtual
corrections and in the cross section for 
real gluon emission, are
treated in two different ways. Both calculations follow subtraction
procedures, one using dimensional regularization and one introducing
an infinitesimal gluon mass \cite{di99}. 
The results
obtained by the two different procedures are in mutual numerical
agreement after adding the contributions from virtual gluon exchange
and real gluon emission. A second, completely independent calculation
of the QCD corrections to the total cross section was based on the
evaluation of all relevant cut diagrams of the photon and $Z$-boson
self-energies in two-loop order, generalizing the method applied to
$t\bar{t}(g)$ intermediate states in Ref.~\cite{dg94}. 
The results of the two approaches are in numerical agreement. Moreover,
the parts that have been calculated in Refs.~\cite{da97,da99},
i.e.~photon exchange and resonant contributions, are in full agreement
with our corresponding partial results.

\section{Results}
{\bf a.) Asymptotic behaviour.}
The QCD corrections to the top final states $t\bar t\phi$ can be interpreted
easily in two kinematical areas. Whenever the invariant mass of the $t\bar
t$ pair is close to 
threshold, the gluonic Sommerfeld rescattering-correction
is positive and becomes large. In the threshold region the $K$
factor approaches the asymptotic form \cite{di98}
\begin{equation}
K^{t\bar t\phi}_{\mathrm{thr}} \to 1+\frac{32\alpha_s}{9\beta_t}
\label{eq:Coulsing}
\end{equation}
with the maximal quark velocity 
$\beta_t=\sqrt{(\sqrt{s}-M_\phi)^2-4 M_t^2}/2M_t$ 
in the $(t\bar t)$ rest frame.

For high energies, on the other hand, the QCD corrections are of order
$\alpha_s/\pi$. In the energy region $s\gg 4M_t^2\gg M_H^2$ but $\log
s/M_t^2 \not \!\gg {\cal O}(1)$, which is relevant for the present
analysis, the QCD corrections can qualitatively
be traced back to vertex corrections and infrared gluon radiation. Since
scalar Higgs bosons are radiated off top quarks preferentially with
small energy [$x=E_\phi/E_t \to 0$], as is evident from the leading
(universal) part of the fragmentation function
\begin{equation}
f(t\to tH;x) = \frac{g^2_{ttH}}{16\pi^2} \left[ 4~\frac{1-x}{x} +
x~\log \frac{s}{M_t^2} \right] \,\, ,
\label{eq:fragh}
\end{equation}
the QCD correction of the scalar Yukawa vertex, regularized by soft
gluon radiation, approaches the value \cite{da97}
\begin{equation}
\Delta^{V+IR}_{H} = \frac{4\alpha_s}{3\pi} \left[ -1+ \frac{2-x}{x}
\log(1-x) \right] \to -4\frac{\alpha_s}{\pi} \,\, .
\end{equation}
The scalar Yukawa vertex is
therefore reduced by four units in $\alpha_s/\pi$ which are compensated
only partly by one unit due to the increase of the $t\bar t$ production
probability, leading in total \cite{di98,da97} to
\begin{equation}
K^{t\bar t\phi}_{\mathrm{cont}} \to 1-3\frac{\alpha_s}{\pi} + \ldots
\hspace*{0.5cm} \mbox{for $\phi=H_{SM},h,H$} \,\, .
\label{eq:kas}
\end{equation}
The ellipsis accounts for hard Higgs and gluon radiation (of order
$+\alpha_s/\pi$). Thus, the QCD corrections are expected negative for
scalar Higgs particles in the high energy continuum.

By contrast, the corresponding fragmentation function for the pseudoscalar
Higgs boson \cite{beenakker}
\begin{equation}
f(t\to tA;x) = \frac{g^2_{ttA}}{16\pi^2} ~x~\log \frac{s}{M_t^2}
\label{eq:fraga}
\end{equation}
is hard so that the average of the vertex and IR gluon corrections over
the Higgs spectrum amounts to
\begin{equation}
\Delta^{V+IR}_A \to \frac{4\alpha_s}{3\pi} \left\langle \left[ 1+ \frac{2-x}{x}
\log(1-x) \right] \right\rangle \sim -\frac{3}{2} \frac{\alpha_s}{\pi} \,\, .
\end{equation}
Adding to this correction the increase of the $t\bar t$ production
probability of one unit, the $K$ factor is very close to unity
\begin{equation}
K^{t\bar tA}_{\mathrm{cont}} \to 1-\frac{1}{2}\frac{\alpha_s}{\pi} + \ldots
\end{equation}
After hard gluon bremsstrahlung is taken into account (symbolized by the
ellipsis), the overall QCD corrections for the pseudoscalar Higgs boson
are therefore expected slightly positive. [For ultra-high energies,
i.e.~$\log s/M_t^2\gg 1$, hard gluon bremsstrahlung becomes important.
Similarly to the leading terms in the fragmentation functions
eqs.~(\ref{eq:fragh}) and (\ref{eq:fraga}), the QCD corrections for
scalar and pseudoscalar Higgs bosons approach each other as a result of
chiral symmetry restoration in {\it asymptotia}; this has been verified
in a numerical calculation.]

Similar estimates can be applied to bottom final states which in general are
dominated by resonance decays. After absorbing the large logarithms
$\log(Q^2_\phi/M_b^2)$ into the Yukawa couplings, the non-leading effects
are positive:
\begin{equation}
K^{b\bar b\phi}_{\mathrm{res}} \approx 1+\left\{{\textstyle \frac{17}{3}},1
\right\} \frac{\alpha_s}{\pi} 
\hspace*{0.5cm}
\mbox{for $\left\{ \mbox{Higgs}, Z\right\} \to b\bar b$ } \,\, .
\end{equation}
Also close to the thresholds and in the high-energy limit the QCD
corrections remain positive after mapping the large (negative) corrections
into the running Yukawa couplings. Since different channels are activated at the
same time, only a qualitative estimate can be given in the continuum
regime,
\begin{equation}
K^{b\bar b\phi}_{\mathrm{cont}} = 1+c\frac{\alpha_s}{\pi} 
\hspace*{0.5cm}
\mbox{with $c = {\cal O}(1)$,}
\end{equation}
while details must be left to the numerical analysis. \\

\begin{figure}[hbt]
\vspace*{0.5cm}

\hspace*{1.5cm}
\begin{turn}{-90}%
\epsfxsize=7cm \epsfbox{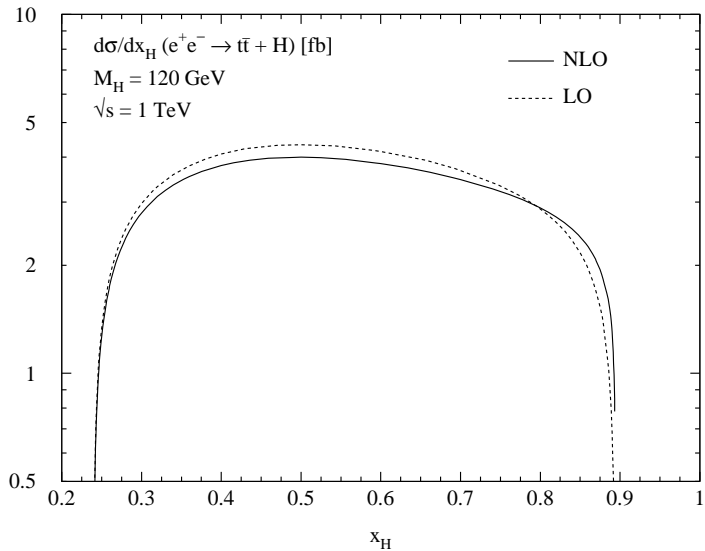}
\end{turn}
\vspace*{-0.2cm}
\caption[ ]{\it
  The Higgs-energy distribution for the process $e^+e^-\to t\bar{t} H + X$,
  including QCD radiative corrections (full curve) and at LO (dashed
  curve) as a function of the scaled Higgs energy $x_H$ \cite{dixx}.}
\label{fg:smxh}
\end{figure}

\noindent
{\bf b.) Numerical results.}
\begin{figure}[hbt]
\vspace*{-1.8cm}

 \epsfysize=10cm
 \epsfxsize=10cm
 \centerline{\epsffile{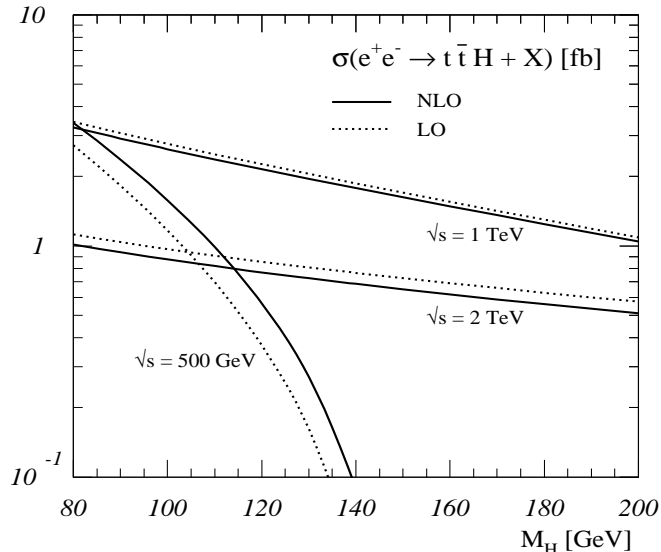}}
\vspace*{-1.0cm}
\caption[ ]{\it
  The total cross section for the process $e^+e^-\to t\bar{t} H + X$,
  including QCD radiative corrections (full curve) and at LO (dashed
  curve) as a function of the scaled Higgs energy $x_H$ \cite{di98}.}
\label{fg:smtot}
\end{figure}
The total cross sections for $Ht\bar t$ production in the SM are
presented in Fig.~\ref{fg:smtot} as a function of the Higgs mass for
different collider energies \cite{di98}. The NLO (LO) cross sections are
given by the full (dotted) curves. The QCD corrections are large
for $\sqrt{s}=500$ GeV due to the Coulomb singularity at threshold and
moderate for high energies. At high energies
the results agree with the approximation of eq.~(\ref{eq:kas}).
The cross sections amount to more than about 0.1 fb, which
leads to a significant number of events at the TESLA collider,
being designed to reach integrated luminosities of about $\int
{\cal L} \sim 1$ ab$^{-1}$ in three years of operation.
The corresponding Higgs-energy distribution is presented in
Fig.~\ref{fg:smxh} for $M_H=120$ GeV as a function of the scaled
variable $x_H=2E_{H}/\sqrt{s}$ \cite{dixx}. The shape of the distribution is
shifted towards larger values of $x_H$ due to the Coulomb singularity at
threshold. The results for small values of $x_H$ reproduce the
approximation of eq.~(\ref{eq:kas}).

\begin{figure}[hbtp]
\vspace*{0.0cm}

\hspace*{-1.7cm}
\begin{turn}{-90}%
\epsfxsize=7cm \epsfbox{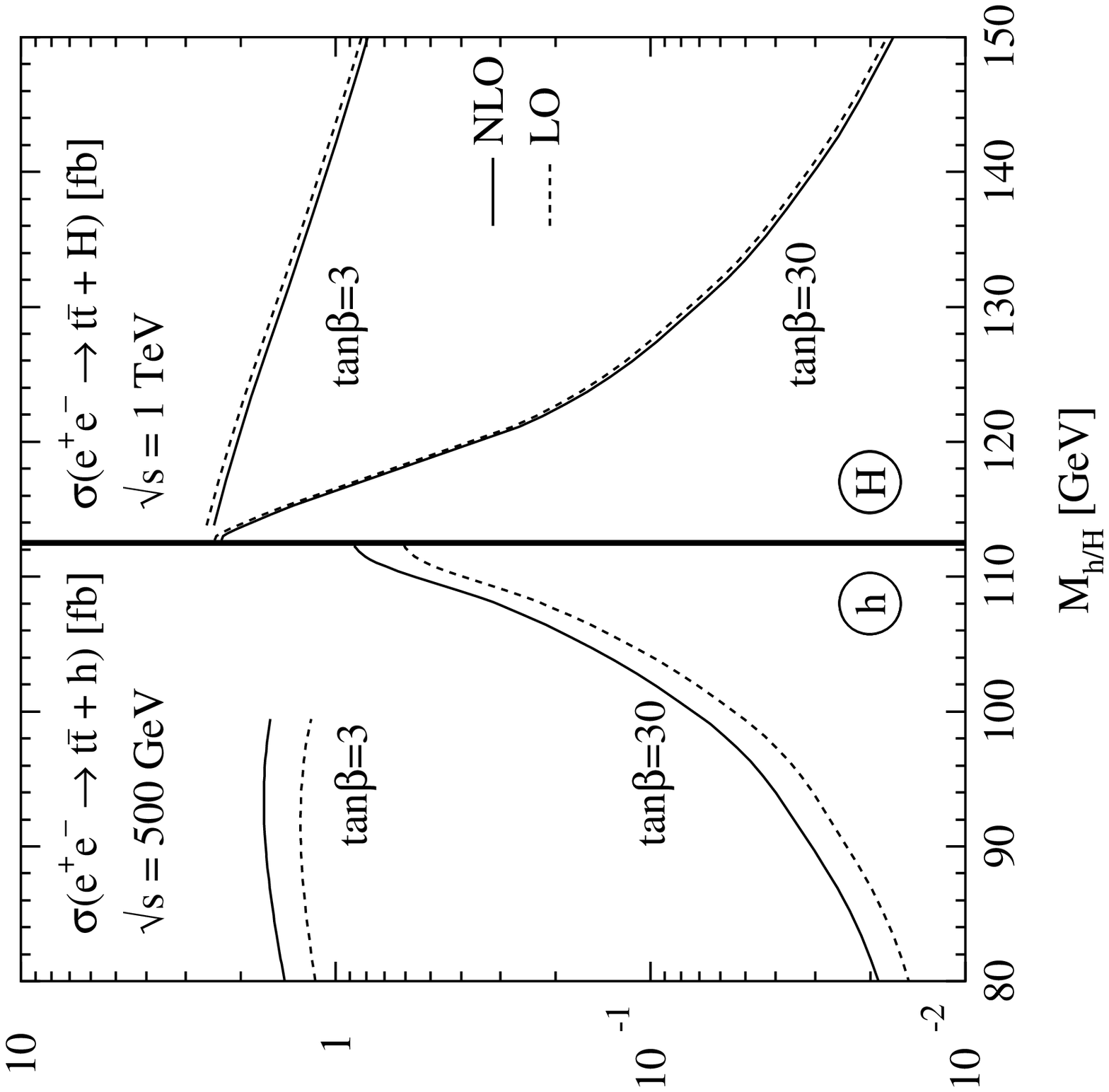}
\end{turn}
\hspace*{-1.8cm}
\begin{turn}{-90}%
\epsfxsize=7cm \epsfbox{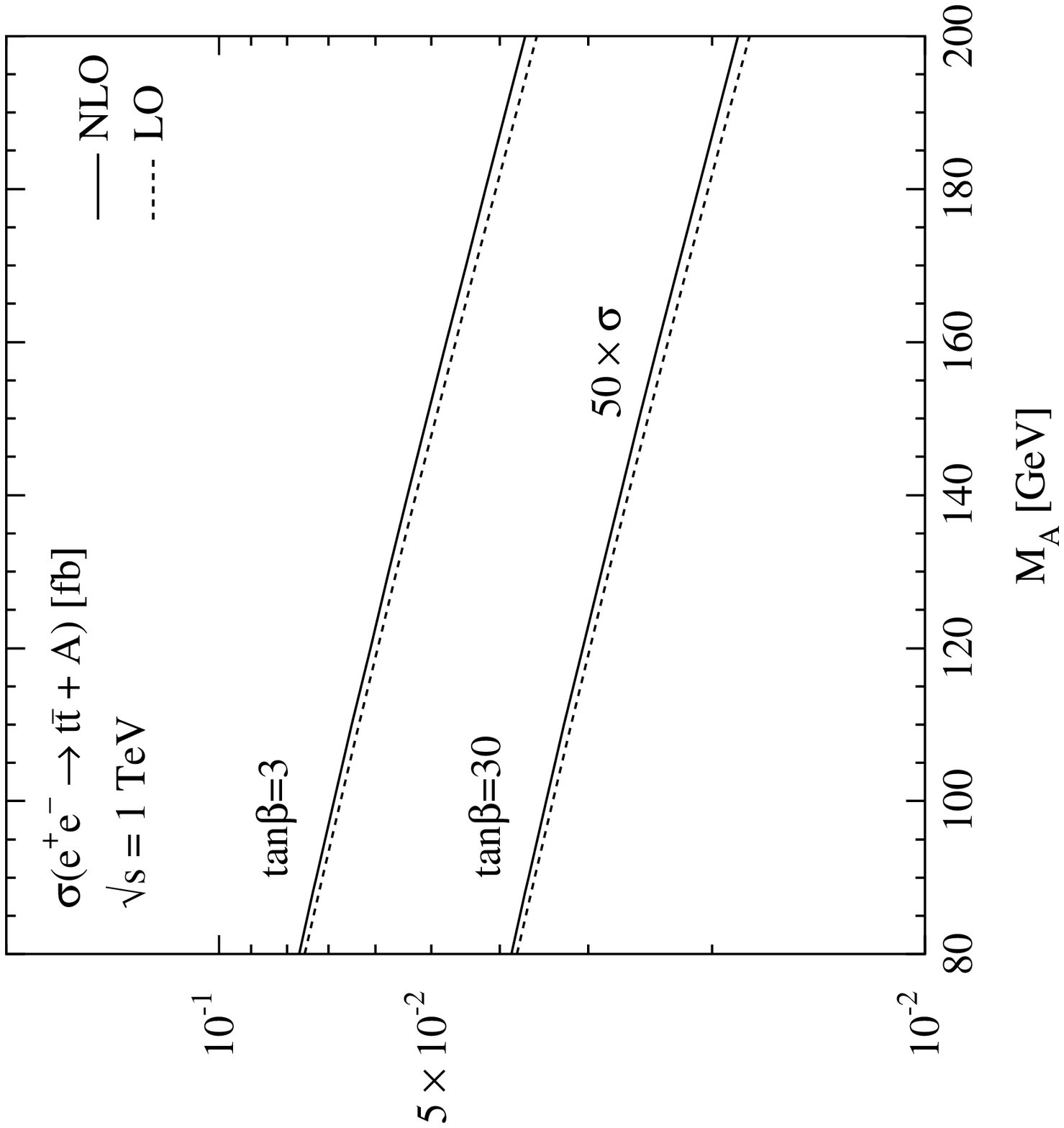}
\end{turn}
\vspace*{0.0cm}

\hspace*{-1.7cm}
\begin{turn}{-90}%
\epsfxsize=7cm \epsfbox{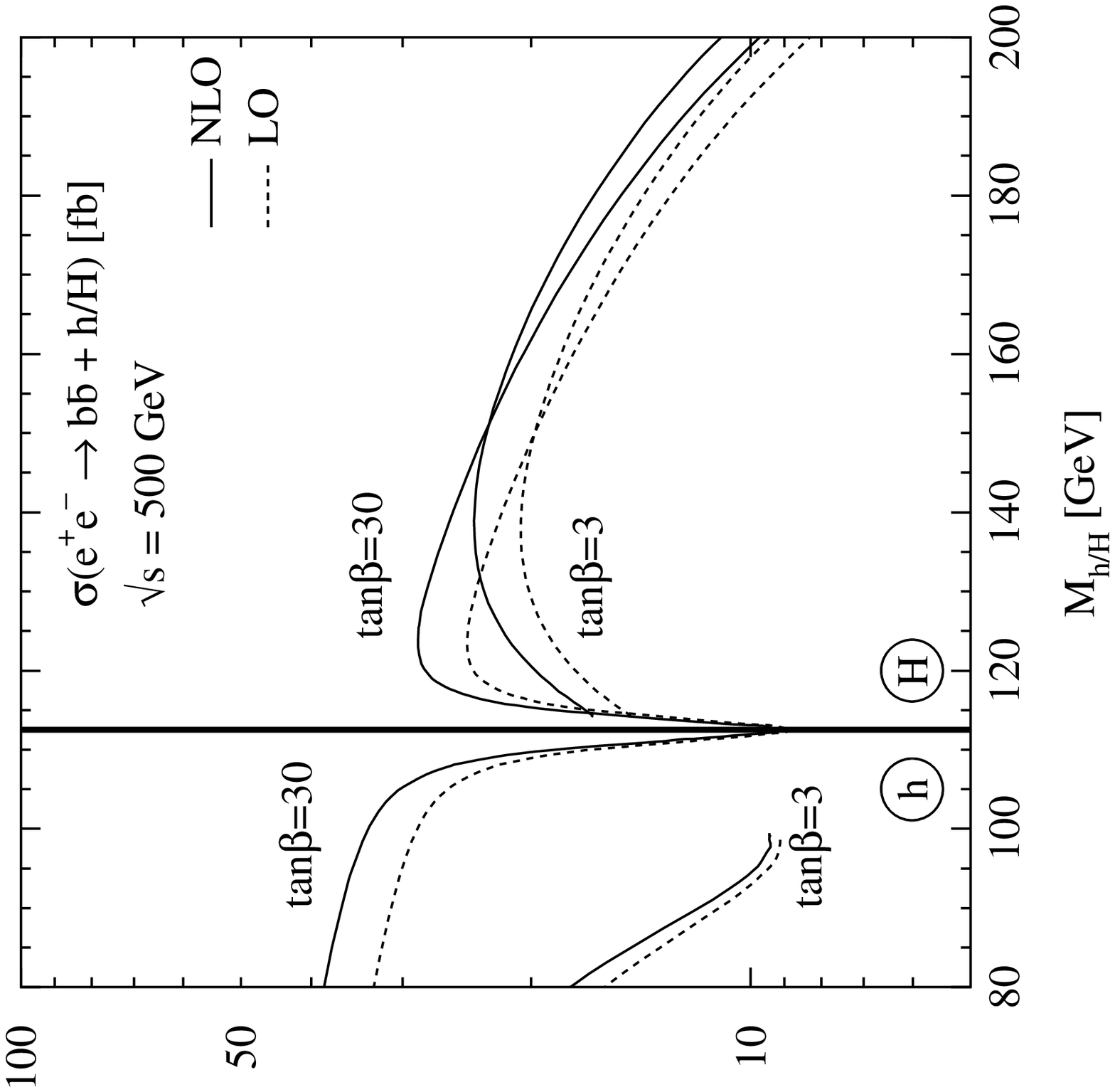}
\end{turn}
\hspace*{-1.8cm}
\begin{turn}{-90}%
\epsfxsize=7cm \epsfbox{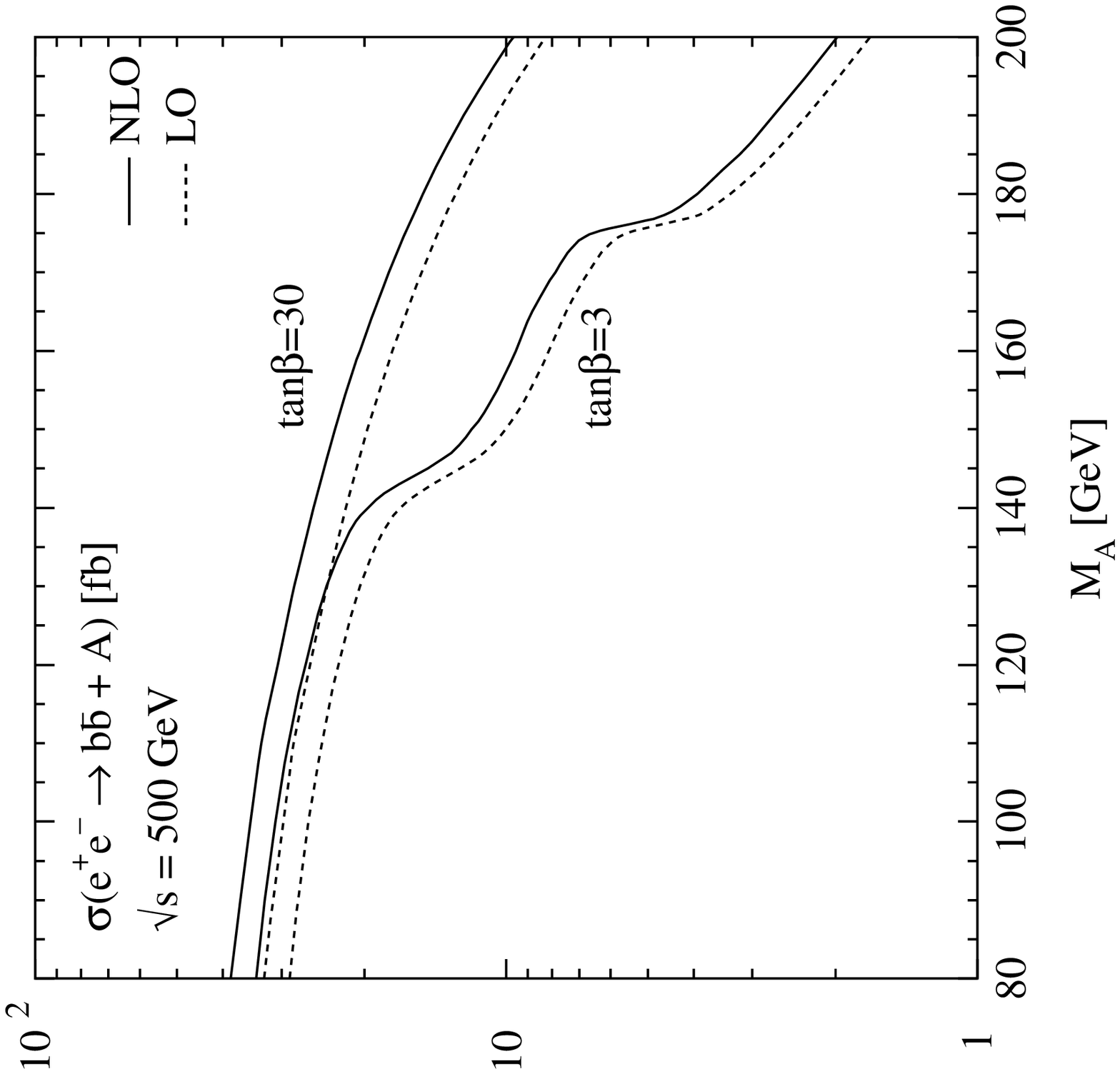}
\end{turn}
\vspace*{-0.5cm}

\caption[ ]{\it 
Production cross sections for the MSSM Higgs bosons $h,H$ and $A$ in
association with heavy $t,b$ quark pairs \cite{di00};
Born approximation: dashed, QCD corrections included: full curves. The rapid
drops in the $b\bar bA$ cross section at $\tgb=3$ are due to the
kinematical opening of the resonant $H\to WW, hh$ decays, which reduce
the branching ratio of the resonant $H\to b\bar b$ decay.
\label{fig:cs}}
\end{figure}

The results for the MSSM are exemplified in Fig.~\ref{fig:cs} for
$\tgb=3$\footnote{Although the value $\tgb=3$ is already excluded for
vanishing mixing in the stop sector, this choice exemplifies the
complete result for smaller values of $\tgb$ and can easily be extended to
the non-vanishing stop mixing case.} and 30 and for the collider energy
$E_{\mathrm{CM}}=500\GeV$ \cite{di00}.
If required by the size of the cross section, which should not fall
below $\sim 10^{-2}$ fb in order to be accessible experimentally, we switched
to $E_{\mathrm{CM}}=1\TeV$.
The Born
terms are shown by the dotted curves, while the final results for the cross
sections, including QCD corrections, are given by the full curves.

For $E_{\mathrm{CM}}=500\GeV$ 
the QCD corrections to Higgs-boson production in
association with $t\bar t$ pairs increase the scalar Higgs-production
cross sections significantly,
as can be inferred from
Fig.~\ref{fig:cs}. Close to threshold the numerical results
clearly exhibit the strong increase of the cross sections due to the
Coulomb singularity (\ref{eq:Coulsing}).
Moreover, for $\tgb=30$ the cross sections
are strongly suppressed except for the regions 
where the light (heavy)
scalar Higgs mass is close to its upper (lower) bound. For
$\tgb=3$ the cross section amounts to about 1 fb, which
leads to a significant number of events at the TESLA collider.

For CM energies of 1 TeV the QCD corrections to scalar Higgs
production in association with $t\bar t$ pairs are of moderate size.
They decrease the cross sections by about 3--5\%. This is in
accordance with the asymptotic form of the $K$ factor \cite{di98}.
For pseudoscalar Higgs production, the size of the QCD corrections is
slightly positive at 1 TeV, in agreement with the qualitative
discussion above.

The cross sections for
Higgs-boson production associated with $b\bar b$
pairs are significantly larger due to the resonance contributions from on-shell
$Z$ and 
Higgs-boson decays into $b\bar b$ pairs. The QCD corrections
increase these cross sections by about 
5--25\%. The drop in the $b\bar bA$ production cross
section for $\tgb=3$ at $M_A\sim 140\GeV$ and $175\GeV$ can be attributed
to the crossing of the thresholds for resonant $H\to WW$ and $H\to hh$
decays, respectively, in $HA$ final states.

\begin{figure}[hbtp]
\vspace*{0.0cm}

\hspace*{-1.7cm}
\begin{turn}{-90}%
\epsfxsize=7cm \epsfbox{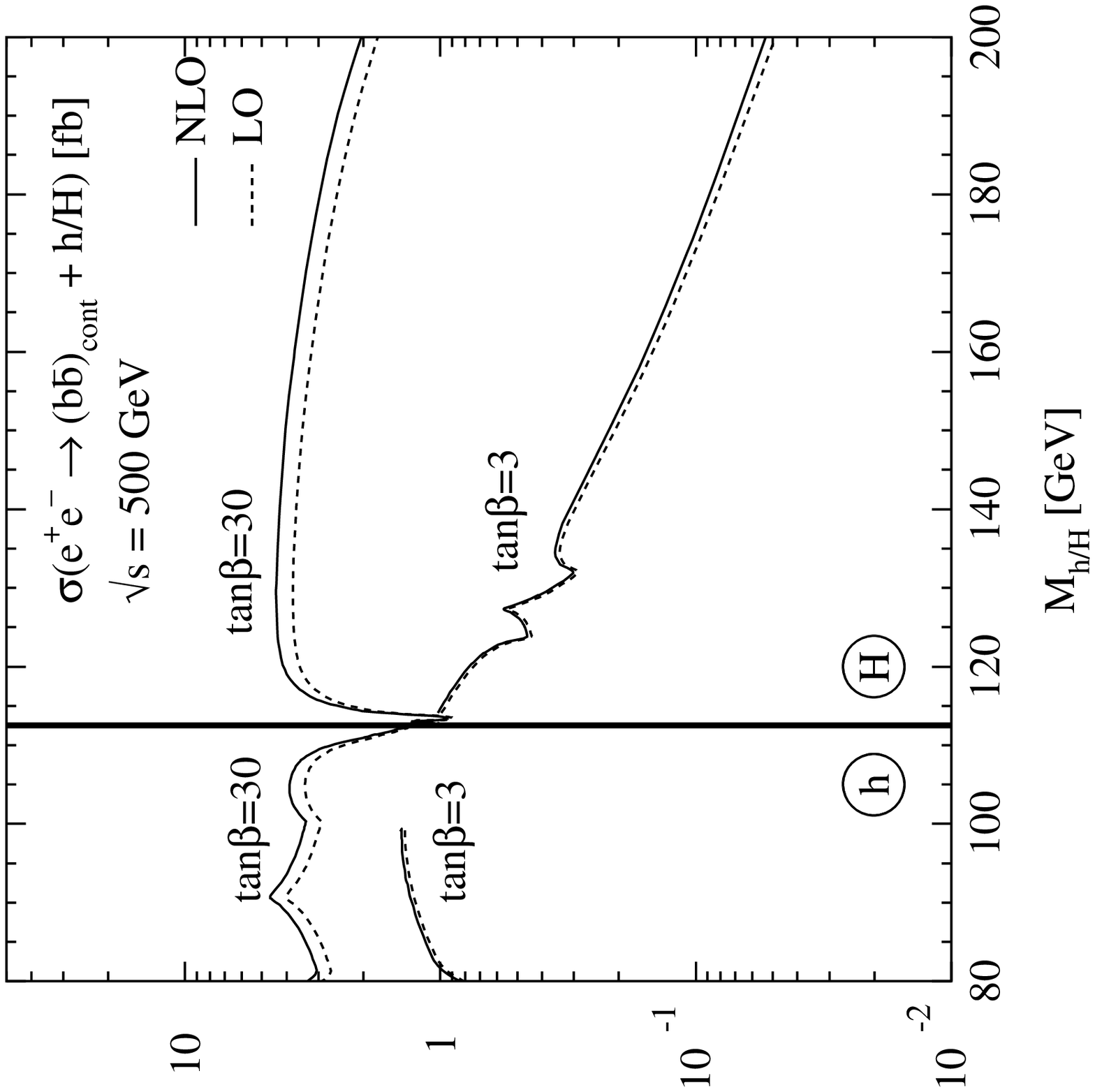}
\end{turn}
\hspace*{-1.8cm}
\begin{turn}{-90}%
\epsfxsize=7cm \epsfbox{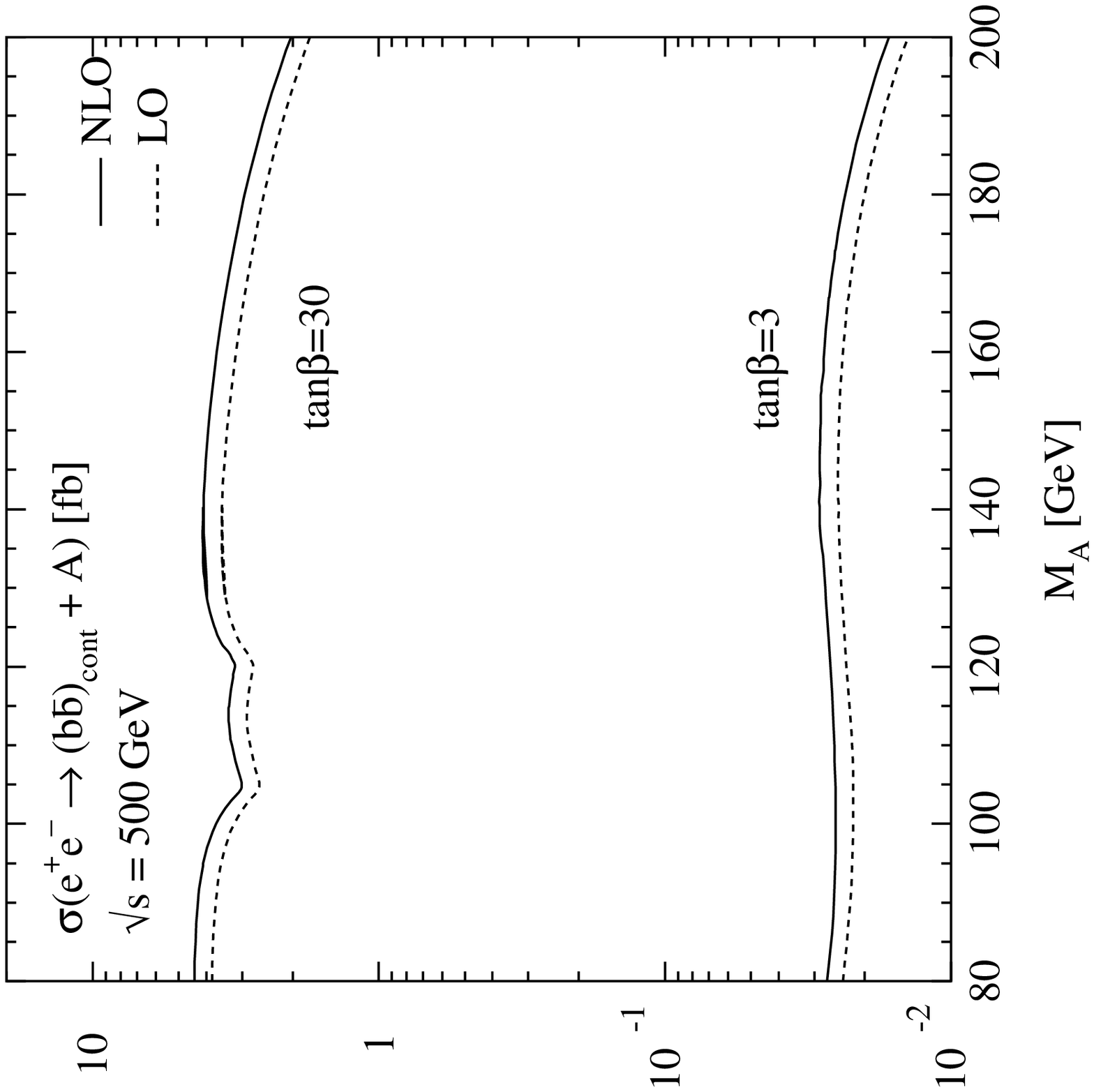}
\end{turn}
\vspace*{-0.5cm}

\caption[ ]{\it Continuum production of MSSM Higgs bosons $h,H,A$ in
association with a $b\bar b$ pair after subtraction of resonance decays
to $b\bar b$ pairs in the Breit-Wigner bands $M_R\pm 5$ GeV \cite{di00}.
\label{fg:cont}}
\end{figure}

Without cuts, 
the intermediate resonance decays $Z,h,H,A\to b\bar b$ dominate all
$b\bar b\phi$ production processes, whenever they are kinematically
allowed, and it will be difficult to extract the bottom Yukawa
couplings experimentally in regions where resonant Higgs decays
to $b\bar b$ pairs are dominant. This is the case at large values of
$\tgb$ for all neutral Higgs particles and at small values of
$\tgb$ for Higgs masses below the $WW$ threshold. In these cases
the branching ratios, which determine the size of the $b\bar b\phi$ 
cross sections, will be nearly independent of the bottom Yukawa couplings.
Resonance decays $R\to b\bar b$ in the $b\bar b h/H/A$ final states can
however be eliminated by cutting out the resonance energy of the final-state
Higgs boson, $E_{\phi,\mathrm{res}} = (s+M^2_\phi-M_R^2)/2\sqrt{s}$.
After subtracting these resonance parts, the non-resonant
contributions are suppressed by about one to three orders of magnitude.
The resonances have been removed from the cross sections in the examples of
Fig.~\ref{fg:cont} by subtracting the two-particle
cross sections in the Breit--Wigner bands $M_R\pm \Delta$ of the energy
$E_{\phi,\mathrm{res}}$ with
the resolution $\Delta = 5\GeV$.
This theoretical definition is used for the sake of simplicity; wider cuts
may be required in experimental analyses.
Peaks and dips in the cross sections are the result of overlapping
Breit-Wigner bands. For the scalar Higgs
particles they arise from overlapping $Z$ and $A$ boson bands; in
pseudoscalar Higgs production the light and heavy scalar
resonance bands overlap for 100 GeV$~\lsim M_A \lsim 120$ GeV for
$\tgb=30$. The dips occur whenever the two resonance bands
touch each other, while the peaks between the dips occur when the
resonance masses coincide exactly. As shown in Fig.~\ref{fg:cont}, the 
QCD-corrected cross sections are still close to $1\fb$ or slightly below,
except for heavy masses at small values of $\tgb$.
Thus, ensembles of order $10^3$ events can be collected at a high-luminosity
collider \cite{di00}.

\begin{figure}[hbtp]
\vspace*{0.0cm}

\hspace*{1.5cm}
\begin{turn}{-90}%
\epsfxsize=7cm \epsfbox{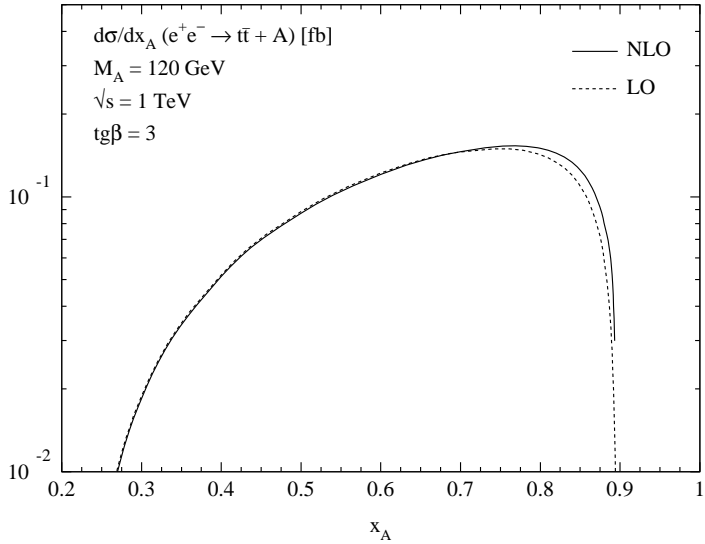}
\end{turn}
\vspace*{-0.2cm}
\caption[ ]{\it
  The Higgs-energy distribution for the process $e^+e^-\to t\bar{t} A + X$,
  including QCD radiative corrections (full curve) and at LO (dashed
  curve) as a function of the scaled Higgs energy $x_A$ for $M_A=120$
  GeV at $\sqrt{s}=1$ TeV \cite{dixx}.}
\label{fg:ttaxa}
\end{figure}

Fig.~\ref{fg:ttaxa} presents the pseudoscalar Higgs-energy distribution in
$t\bar tA$ production at $\sqrt{s}=1$ TeV for $M_A=120$ GeV and $\tgb=3$
as a function of the scaled Higgs energy $x_A=2E_A/\sqrt{s}$ \cite{dixx}.
As in the
case of the SM Higgs boson the distribution is shifted towards larger
Higgs energies due to the Coulomb singularity at threshold. Moreover, it
is clearly visible from the comparison of Figs.~\ref{fg:smxh} and
\ref{fg:ttaxa} that the pseudoscalar energy spectrum is harder than the
scalar one, which is related to the absence of the $1/x$ enhancement in
eq.~(\ref{eq:fraga}).

\begin{figure}[hbtp]
\vspace*{0.0cm}

\hspace*{1.5cm}
\begin{turn}{-90}%
\epsfxsize=7cm \epsfbox{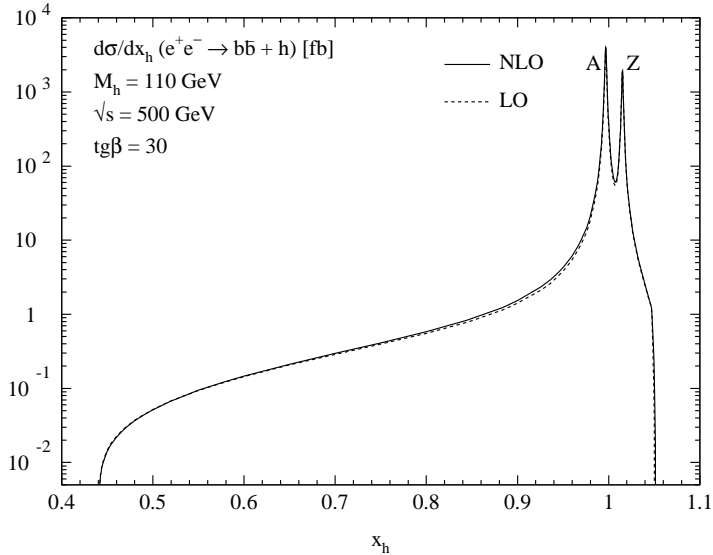}
\end{turn}
\vspace*{-0.2cm}
\caption[ ]{\it
  The Higgs-energy distribution for the process $e^+e^-\to b\bar{b} h + X$,
  including QCD radiative corrections (full curve) and at LO (dashed
  curve) as a function of the scaled Higgs energy $x_h$ for $M_h=110$ 
  GeV at $\sqrt{s}=500$ GeV \cite{dixx}.}
\label{fg:bbhxh}
\end{figure}

\begin{figure}[hbtp]
\vspace*{0.0cm}

\hspace*{1.5cm}
\begin{turn}{-90}%
\epsfxsize=7cm \epsfbox{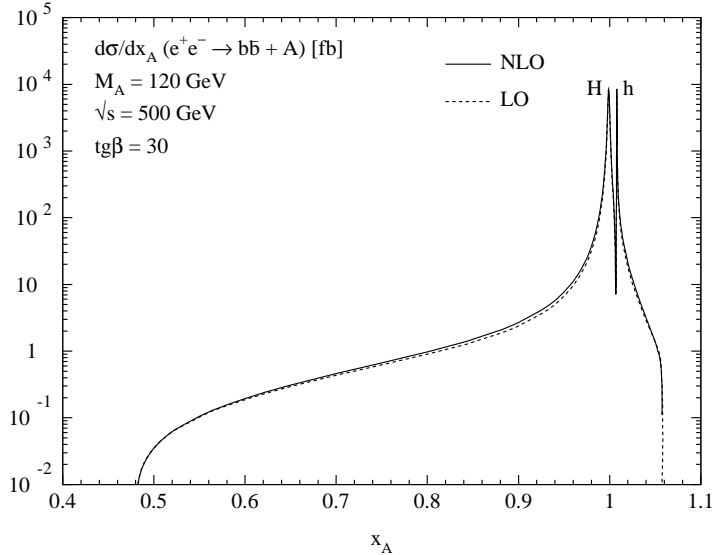}
\end{turn}
\vspace*{-0.2cm}
\caption[ ]{\it
  The Higgs-energy distribution for the process $e^+e^-\to b\bar{b} A + X$,
  including QCD radiative corrections (full curve) and at LO (dashed
  curve) as a function of the scaled Higgs energy $x_A$ for $M_A=120$ 
  GeV at $\sqrt{s}=500$ GeV \cite{dixx}.}
\label{fg:bbaxa}
\end{figure}

The corresponding Higgs-energy distributions of $b\bar bh$ production for
$M_h=110$ GeV is exemplified in Fig.~\ref{fg:bbhxh} and for $b\bar bA$
production for $M_A=120$ GeV in Fig.~\ref{fg:bbaxa} as functions of the
corresponding scaled Higgs energies for $\tgb=30$ at $\sqrt{s}=500$ GeV
\cite{dixx}.
The resonance peaks of
$A,Z\to b\bar b$ are dominating the distribution of $b\bar bh$ production
and of $h,H\to b\bar b$ the distribution of $b\bar bA$ production. The
continuum contributions are suppressed by several orders of magnitude. The
QCD corrections are moderate for all Higgs energies.

\section{Conclusions}
Top- and bottom-Yukawa couplings in the SM and supersymmetric theories
can be measured at future linear
$e^+e^-$ colliders in Higgs radiation off top and bottom quarks. We have
presented the total cross sections and Higgs-energy distributions
including the full NLO QCD corrections. The corrections turn out to be
large for Higgs radiation off top quarks at c.m.~energies $\sqrt{s}=500$
GeV, while they are moderate at larger energies. The QCD corrections to
Higgs radiation off bottom quarks are moderate after absorbing large
logarithms in the running bottom Yukawa couplings.

Measurements of the top Yukawa coupling in SM Higgs radiation off top
quarks can be performed quite accurately at the future TESLA collider
thanks to the high luminosities.
Measurements of Yukawa couplings in supersymmetric Higgs radiation
off heavy quarks at $e^+e^-$ linear colliders are difficult. This is a
result of the large number of subchannels contributing to the $Q\bar Qh/H/A$
final states in supersymmetric theories in general, and the contamination by
two-Higgs final states in particular. Nevertheless, the continuum cross
sections appear large enough to allow for a solution of this experimental
problem as
shown in the present analysis. Even though experimental simulations are beyond
the scope of this note, it may be concluded from earlier SM analyses that the
method will work at least in part of the MSSM parameter space, thus providing us
with the absolute size of the quark--Higgs Yukawa couplings in the minimal
supersymmetric theory. \\[0.5cm]

\noindent
{\bf Acknowledgements.} I would like to thank S.~Dittmaier, M.~Kr\"amer,
Y.~Liao and P.M.~Zerwas for the pleasant collaboration on the topic
presented here. Moreover, I am grateful to the organizers of
RADCOR--2000 for the stimulating atmosphere during the symposium and
financial support.

\end{document}